\documentstyle[12pt]{article}
\newcommand{\bi}{\bibitem}
\newcommand{\Vbf}{\mbox{\boldmath $V$}}
\newcommand{\ubf}{\mbox{\boldmath $u$}}
\newcommand{\pbof}{\mbox{\boldmath $p$}}
\newcommand{\Le}{{\cal L}}

\newcommand{\longr}{\longrightarrow}

\newcommand{\infi}{\infty}

\newcommand{\De}{\Delta}

\newcommand{\bb}{\begin{equation}}
\newcommand{\ee}{\end{equation}}
\newcommand{\bega}{\begin{eqnarray}}
\newcommand{\ega}{\end{eqnarray}}
\newcommand{\begae}{\begin{eqnarray*}}
\newcommand{\egae}{\end{eqnarray*}}

\newcommand{\up}{\uparrow}
\newcommand{\dow}{\downarrow}

\newcommand{\h}{\hspace*{4ex}}

\newcommand{\ov}{\overline}

\newcommand{\cent}{\centerline}
\newcommand{\vs}{\vspace*}

\begin{document}

\baselineskip 0.65cm

\begin{center}

{\large {\bf The Tolman ``Antitelephone'' Paradox: Its Solution by\\
Tachyon Mechanics}$^{\: (\ast)}$}
\footnotetext{$^{\: (\ast)}$  Work partially supported by CNPq (Brazil), and by
INFN--Sezione di Catania, MURST and CNR (Italy).}

\end{center}

\vs{5mm}

\cent{ Erasmo RECAMI }

\vs{0.5 cm}

\cent{{\em Facolt\`a di Ingegneria, Universit\`a Statale di Bergamo, Dalmine
(BG), Italy;}}

\cent{{\em INFN--Sezione di Milano, Milan, Italy; \ and}}

\cent{{\em Dept. Applied Mathematics, State University of Campinas,
Campinas, S.P., Brazil.}}

\vs{2. cm}

{\bf Abstract  \ --} \ Some recent experiments led to the claim that
something can travel faster than light in vacuum. \ However, such results
do not seem to place relativistic causality in jeopardy. \ Actually, it is
possible to solve also the known causal paradoxes, devised for ``faster
than $c \:$" motion: even if this is not widely recognized. \ Here we want
to show, in detail
and rigorously, how to solve the oldest causal paradox, originally proposed by
Tolman, which is the kernel of so many further tachyon paradoxes. The key to
the
solution  is a careful application of {\em tachyon mechanics}, that can be
unambiguously derived from special relativity.

\


\vfill\eject

\h {\em Introduction.} --  Some recent experiments, performed at
Cologne($^{1}$), Berkeley($^{2}$) and Florence($^{3}$) led to the claim
that {\em something} can travel with a speed larger than
the speed $c$ of light in vacuum, thus confirming some older
predictions.($^{4}$)
 \ Nevertheless, such results do not seem to
place relativistic causality in jeopardy.

\h Actually, it is possible
to solve also the known causal paradoxes, devised for ``faster
than light" motion, even if this is not widely recognized.

\h In fact, claims exist since long($^{5}$)
that all the ordinary causal paradoxes proposed for tachyons can be solved
(at least ``in microphysics"($^{6}$))
on the basis of the ``switching procedure" (swp) introduced
by St\"{u}ckelberg($^{7}$), Feynman($^{7}$) and Sudarshan($^{5}$), also
known as the reinterpretation principle: a principle which in refs.($^{8,9}$)
has been given the status of a fundamental postulate
of special relativity (SR), both for bradyons [slower--than--light
particles] and for tachyons. \ Schwartz,($^{10}$) at last, gave the swp
a formalization in which it becomes ``automatic".

\h However, the effectiveness  of the swp and of those solutions is often
overlooked, or misunderstood. \
Here we want therefore to show, in detail and rigorously,
how to solve the oldest ``paradox", i.e. the {\em antitelephone} one,
originally proposed by Tolman($^{11}$) and then reproposed by many
authors. We shall refer to its recent formulation  by Regge,($^{12}$)
 and spend some
care in solving it, since it is the kernel of many other paradoxes.
Let us stress that: \ (i) any careful solution of the tachyon causal
``paradoxes" has to make recourse to explicit calculations based on the
mechanics of tachyons; \ (ii) such tachyon mechanics can be
unambiguously and univocally derived from SR, by referring the
Superluminal objects to the class of the ordinary, subluminal observers
{\em only} (i.e., without any need of introducing ``Superluminal
reference frames''); \ (iii) moreover, the comprehension of the whole
subject will be
substantially enhanced if one will refer himself to
the (subluminal, ordinary) SR based on
the {\em whole} proper Lorentz group $\Le_+ \equiv \Le^\up_+ \cup
\Le^\dow_+$, rather than on its orthochronous subgroup $\Le^\up_+$ only
[see refs.($^{9}$), and references therein]. \ At last, for a modern
approach to the classical theory of tachyons, reference can be made to
the review article($^{6}$) as well as to refs.($^{13,14}$).\\

\h {\em Tachyon mechanics.} -- In ref.($^{15}$) the basic tachyon mechanics
can be found exploited for the processes: \ a) proper (or
``intrinsic") emission of a tachyon T by an ordinary body A; \ b)
``intrinsic" absorption of a tachyon T by an ordinary body A; \ c)
exchange of a tachyon T between two ordinary bodies A and B. \
The word ``intrinsic" refers to the fact that those processes
(emission, absorption by A) are described {\em as they appear}
in the rest-frame of
A; particle T can represent both a tachyon and an antitachyon.
Let us recall the following results only.

\h Let us first consider a tachyon moving with velocity $\Vbf$ in
a reference frame $s_0$. If we pass to a second frame $s'$, endowed with
velocity $\ubf$ w.r.t. (with respect to) frame $s_0$, then the new
observer $s'$ will see ---instead of the initial tachyon T--- an
antitachyon $\ov{\rm T}$ travelling the opposite way in space (due to the
swp), if and only if\\

(1) {\hfill{
$ \ubf \cdot \Vbf > c^2 \; . $
\hfill}}

\

Recall in particular that, if $\ubf \cdot \Vbf < 0$, the ``switching"
does {\em never} come into play.

\h Now, let us explore some of the unusual and unexpected
consequences of the trivial fact that in the case of tachyons it is\\

(2) {\hfill{
$ |E| = + \sqrt{\pbof^2-m^2_0} \qquad (m_0 \ \ \mbox{real}; \ {\Vbf}^2 > 1)
\; , $
\hfill}}

\

where we chose units so that, numerically, $\: c=1$.

\h Let us, e.g., describe the phenomenon of ``intrinsic emission" of
a tachyon, as seen in the rest frame of the emitting body: Namely,
let us consider {\em in its rest frame} an ordinary body A, with
initial rest mass $M$, which emits a tachyon (or antitachyon) T
endowed with (real) rest mass $m \equiv m_0$, four-momentum $p^\mu
\equiv (E_{\rm T}, \pbof)$, and velocity $\Vbf$ along the $x$-axis.
Let $M'$ be the
final rest mass of body A. The four-momentum conservation requires\\

(3) {\hfill{
$M= \sqrt{\pbof^2-m^2} + \sqrt{\pbof^2+ M'^{\, 2}}$
\hfill}} (rest frame)\\

that is to say [$V \equiv |\Vbf|$]:\\

(4) {\hfill{
$ 2M |\pbof| = [(m^2+ \De)^2 + 4m^2 M^2]^{\frac{1}{2}} \ ; \quad V= [1+
4m^2 M^2/ (m+\De)^2]^{\frac{1}{2}} \; , $
\hfill}}

\

where [calling  $E_{\rm T} \equiv + \sqrt{\pbof^2-m^2} \,$]:\\

(5) {\hfill{
$\De\equiv M'^{\, 2} - M^2= -m^2- 2ME_{\rm T} \; ,$
\hfill}} (emission)\\

so that\\

(6) {\hfill{
$-M^2 < \De \leq - |\pbof|^2 \leq - m^2 \; .$
\hfill}} (emission)\\

It is essential to notice that $\De$ is, of course, an {\em
invariant} quantity, which in a generic frame $s$ writes\\

(7) {\hfill{
$ \De= -m^2 - 2p_\mu P^\mu \; , $
\hfill}}

\

where $P^\mu$ is the initial four-momentum of body A w.r.t. frame
$s$.

\h Notice that in the generic frame $s$ the process of (intrinsic)
emission can appear either as a T emission or as a $\ov{\rm T}$ absorption
(due to a possible ``switching") by body A. \ The following theorem, however,
holds:($^{15}$)\\

{\em Theorem 1:} $<<$ A necessary and sufficient condition for a
process to be a tachyon emission in the A rest-frame (i.e., to be
an {\em intrinsic emission}) is that during the process the body A
{\em lowers} its rest-mass (invariant statement!) in such a way that
$-M^2 < \De \leq -m^2$.~$>>$\\

\h Let us now describe the process of ``intrinsic absorption" of a
tachyon by body A; \ i.e., let us consider an ordinary body A to
absorb {\em in its rest} frame a tachyon (or antitachyon) T,
travelling again with speed $V$ along the $x$-direction. The
four-momentum conservation now requires\\

(8) {\hfill{
$M+ \sqrt{\pbof^2-m^2} = \sqrt{\pbof^2+ M'^{\, 2}} \; ,$
\hfill}} (rest frame)\\

which corresponds to\\

(9) {\hfill{
$\De \equiv M'^{\, 2} - M^2= -m^2+ 2 M E_{\rm T}  \; ,$
\hfill}} (absorption)\\

so that\\

(10) {\hfill{
$-m^2 \leq \De \leq + \infi \; .$
\hfill}} (absorption)\\

In a generic frame $s$, the quantity $\De$ takes the invariant form\\

(11) {\hfill{
$ \De= -m^2+ 2 p_\mu P^\mu \; . $
\hfill}}

\

It results in the following new theorem:\\

{\em Theorem 2:} $<<$ A necessary and sufficient condition for a
process (observed either as the emission or as the absorption of a
tachyon T by an ordinary body A) to be a tachyon absorption in
the A-rest-frame  ---i.e., to be an {\em intrinsic absorption}---  is
that $\De \geq - m^2$.~$>>$\\

We now have to describe the {\em tachyon exchange} between two
ordinary bodies A and B. We have to consider the four-momentum
conservation at A {\em and} at B; we need to choose a (single) frame
relative to which we describe the whole interaction; let us choose the
rest-frame of A. Let us explicitly remark, {\em however}, that  ---when
bodies A and B exchange one tachyon T---  the tachyon mechanics
is such that the ``intrinsic descriptions" of the processes at A
{\em and} at B can a priori correspond to one of the following four
cases($^{15}$):

\

\setcounter{equation}{11}
\bb
\left\{\begin{array}{ll}
1) & \quad\mbox{emission---absorption} \ ,\\
\\
2) & \quad\mbox{absorption---emission} \ ,\\
\\
3) & \quad\mbox{emission---emission} \ ,\\
\\
4) & \quad\mbox{absorption---absorption} \ .
\end{array}\right.
\ee

\

Case 3) can happen, of course, only when the tachyon exchange takes
place in the receding phase (i.e., while A, B are receding from
each other); case 4) can happen, by contrast, only in the
approaching phase.

\h Let us consider here only the particular tachyon exchanges in
which we have an ``intrinsic emission" at A, and in which moreover the
velocities $\ubf$ of B and $\Vbf$ of T w.r.t. body A are such that $\ubf
\cdot \Vbf > 1$. \ Because of the last condition and the consequent
``switching" (cf. Eq.(1)), from the rest-frame of B one will therefore
observe the flight of an antitachyon $\ov{\rm T}$ {\em emitted} by B and
absorbed by A \ (the {\em necessary} condition for this to happen, let us
recall, being  that A, B  {\em recede} from each other).

\h More generally, the kinematical conditions for a tachyon to be
exchangeable between A and B can be shown to be the
following:\\

I) \ Case of ``intrinsic emission" at A:
\bb
\left\{\begin{array}{l}
 \ \mbox{if} \ \ubf \cdot \Vbf < 1 \ , \quad\mbox{then} \ \De_{\rm B} > -
m^2 \quad (\longr \mbox{intrinsic absorption at B});\\
\\
 \ \mbox{if} \ \ubf \cdot \Vbf > 1 \ , \quad\mbox{then} \ \De_{\rm B} < -
m^2 \quad (\longr \mbox{intrinsic emission at B}).\\
\end{array}\right.
\ee

II) \ Case of ``intrinsic absorption" at A:
\bb
\left\{\begin{array}{l}
 \ \mbox{if} \ \ubf \cdot \Vbf < 1 \ , \quad\mbox{then} \ \De_{\rm B} < -
m^2 \quad (\longr \mbox{intrinsic emission at B});\\
\\
 \ \mbox{if} \ \ubf \cdot \Vbf > 1 \ , \quad\mbox{then} \ \De_{\rm B} > -
m^2 \quad (\longr \mbox{intrinsic absorption at B}).\\
\end{array}\right.
\ee

\h Now, let us finally pass to examine the Tolman paradox.\\

\h {\em The paradox.} -- In Figs.1,2 the axes $t$ and $t'$ are the
world-lines of two devices A and B, respectively, which are able to
exchange tachyons and move with constant relative speed $u$, [$u^2 <
1$], along the $x$-axis. According to the terms of the paradox (Fig.1),
device A sends tachyon 1 to B (in other words, tachyon 1 is
supposed to move forward in time w.r.t. device A). The device B
is constructed so as to send back tachyon 2 to A as soon as it
receives tachyon 1 from A. If B has to {\em emit} (in its
rest-frame) tachyon 2, then 2 must move forward in time w.r.t. device
B; that is to say, the world-line ${\rm BA}_2$  must have a slope {\em
lower} than the slope ${\rm BA}'$ of the $x'$-axis (where ${\rm BA}' // x'$):
 \ this means that ${\rm A}_2$ must stay above ${\rm A}'$. If the speed of
tachyon 2 is
such that ${\rm A}_2$ falls between ${\rm A}'$ and ${\rm A}_1$, it seems
that 2 reaches A (event ${\rm A}_2$) {\em before} the emission of 1
(event ${\rm A}_1$).
This appears to realize an {\em anti-telephone}.\\

\h {\em The solution.} -- First of all, since tachyon 2 moves
backwards in time w.r.t. body A, the event ${\rm A}_2$ will appear to
A as the emission of an antitachyon $\ov{2}$. \ The observer ``$\: t \:$"
will see his own device A (able to exchange tachyons) emit
successively towards B the antitachyon $\ov{2}$ and the tachyon 1.

\h At this point, some supporters of the paradox (overlooking tachyon
mechanics, as well as relations (12)) would say that, well, the
description put forth by the observer ``$\: t \:$" can be orthodox, but
then the device B is no longer working according to the stated programme,
because B is no longer emitting a tachyon 2 on receipt of tachyon
1. \ Such a statement would be wrong, however, since the fact that
``$\: t \:$" observes an ``intrinsic emission" at ${\rm A}_2$ {\em does not
mean} that ``$\: t' \:$" will see an ``intrinsic absorption" at B! \ On the
contrary, we are just in the case considered above, between eqs. (12)
and (13): intrinsic emission by A, at ${\rm A}_2$, with $\ubf \cdot
\Vbf_{\ov{2}} > c^2$, where $\ubf$ and $\Vbf_{\ov{2}}$ are the velocities of
B and $\ov{2}$ w.r.t. body A, respectively; so that {\em both}
A {\em and} B experience an intrinsic {\em emission} (of tachyon 2 or of
antitachyon $\ov{2}$) in their own rest frame.

\h But the tacit premises underlying the  ``paradox" (and even the very
terms in which it was formulated) were ``cheating" us
{\em ab initio}. \ In fact, Fig.1 makes it clear that, if $\ubf \cdot
\Vbf_{\ov{2}} > c^2$, then for tachyon 1 {\em a fortiori} $\ubf \cdot \Vbf_1 >
c^2$, where $\ubf$ and $\Vbf_1$ are the velocities of B and 1 w.r.t. body
A. \ Therefore, due to the previous consequences of tachyon mechanics,
observer
``$t' \,$" will see B intrinsically {\em emit} also tachyon 1 (or, rather,
antitachyon $\ov{1}$). \ In conclusion, the proposed chain of events does
{\em not} include any tachyon absorption by B (in its rest frame).

\h For body B to {\em absorb} (in its own rest frame) tachyon 1,
the world-line of 1 ought to have a slope {\em higher} than the slope
of the $x'$-axis (see Fig.2). Moreover, for body B to {\em emit}
(``intrinsically") tachyon 2, the slope of the of 2 should be lower
than the $x'$-axis'. In other words, when the body B, programmed to
emit 2 as soon as it receives 1, does actually do so, the event ${\rm A}_2$
does happen {\em after} ${\rm A}_1$ (cf. Fig.2), as requested by
causality.\\

\h {\em The moral.} -- The moral of the story is twofold: \ i) one
should never {\em mix} the descriptions (of one phenomenon)
yielded by different observers; otherwise ---even in ordinary
physics--- one would  immediately meet contradictions: in Fig.1, e.g., the
motion direction of 1 is assigned by A and the motion-direction of
2 is assigned by B; this is ``illegal"; \ ii) when proposing a problem
about tachyons, one must comply($^{5}$) with the rules of tachyon
mechanics($^{15}$); this is analogous to complying with the laws of
{\em ordinary} physics when formulating the text of an {\em ordinary}
problem (otherwise
the problem in itself will be ``wrong").

\h Most of the paradoxes proposed in the literature suffered the
above shortcomings.

\h Notice once more that, in the case of Fig.1, neither A nor B regard
event ${\rm A}_1$ as the cause of event ${\rm A}_2$ (or {\em vice-versa}).
In the case of Fig.2, on the other hand, both A and B consider event
${\rm A}_1$ to be the cause of event ${\rm A}_2$: but in this case
${\rm A}_1$ does
chronologically precede ${\rm A}_2$ according to both observers, in
agreement with the relativistic covariance of the law of retarded
causality.

\vs{0.5 cm}

\h The author gladly acknowledges stimulating discussions with
A.O. Barut, H.C. Corben, A. Gigli, P.-O. L\"{o}wdin, M. Jammer, R. Mignani,
E.C.G. Sudarshan and Sir Denys Wilkinson; and thanks Professor U. Gerlach
for a careful reading of the manuscript.

\vs{5. cm}

{\bf Figure captions:}

\

{\bf Fig.1} -- The apparent chain of events, according to the terms of the
paradox.

\

{\bf Fig.2} -- Solution of the paradox: see the text.

\end{document}